\documentclass{PoS}

\usepackage{bm}
\usepackage{amsmath}
\usepackage{yfonts}

\usepackage{graphicx}
\usepackage{rotating}
\usepackage{latexsym}

\newcommand{\beq}{\begin{equation}}
\newcommand{\eeq}{\end{equation}}
\newcommand{\bqa}{\begin{eqnarray}}
\newcommand{\eqa}{\end{eqnarray}}
\newcommand{\qh}{{\hat q}_y}

\def\slashh{\hskip-0.5em/ }

\title{The transport coefficient $\hat{q}$  in an anisotropic plasma}

\ShortTitle{The transport coefficient  $\hat{q}$ in an anisotropic plasma}

\author{Yacine Mehtar-Tani\\
       Institut f\"ur Theoretische Physik, Universit\"at Heidelberg,
 D-69120 Heidelberg, Germany\\
        E-mail: \email{mehtar@thphys.uni-heidelberg.de}}

\abstract{We investigate the jet quenching parameter in the case of a fast moving 
quark in an anisotropic plasma. In the leading log approximation, strong 
indications are found that the transport coefficient increases with 
increasing anisotropy. Implications for the phenomenology at RHIC are 
discussed.

}

\FullConference{High-pT Physics at LHC -09\\
		 February 4- 4 2009\\
		 Prague, Czech Republic}

\begin{document}

\section{Introduction}

In this talk, which is based on the 
work \cite{Baier:2008js}, the transport coefficient in an anisotropic 
but homogenous plasma at high temperature is discussed.
The quoted paper contains a full list of references.

The observation at RHIC of a strong suppression of large momentum particles in
 Au-Au collisions
 when compared to proton-proton collisions scaled with the number of
 participant nucleons,
 is certainly a strong evidence for the formation of a dense partonic
 medium, as e.g. reviewed recently in \cite{Tannenbaum:2006ch}. 

Jet quenching is usually attributed to radiative parton energy loss in the
 medium (for reviews, see: \cite{Baier:2000mf,
Kovner:2003zj,Salgado:2005zp,CasalderreySolana:2007pr}),
 where the properties of the
 medium are
 encoded in the transport coefficient $\hat{q}$ 
that is defined as the ratio of the mean $p_t^2$, transfered from the plasma 
to the hard partons propagating through the medium and the hard parton path
 length
 in the medium. Therefore, the jet quenching parameter is also related
 to the $p_t$-broadening of the energetic parton. More precisely, it is given by 
\beq
\hat{q} = \rho \int d^2q_\perp q^2_\perp \frac{d \sigma}{d^2q_\perp} ~,
\label{eq:basic}
\eeq
where $\rho$ is the number density of the constituents of the medium,
and $\frac{d \sigma}{d^2q_\perp}$ is the differential  scattering cross section of
the parton (massless quark or gluon) on the medium. 
Depending on the model used in phenomenological works and because of the many 
 theoretical uncertainties,  $\hat q$ may be quoted in wide range of
 $0.5-20~GeV^2/fm$ 
see e.g \cite{Eskola:2004cr,Baier:2006fr}.   Therefore,
 the information about the nature of the medium contained in the transport
 coefficient
 is still uncertain. In these studies, so far, the medium is assumed to be isotropic. 

It has been found (see for a recent review:
\cite{Mrowczynski:2005ki,Strickland:2006pb} and e.g.  \cite{Romatschke:2003ms,Romatschke:2003vc,
Romatschke:2004jh})
that the physics of anisotropic plasmas
 differs from that of isotropic
ones, because of the presence of plasma instabilities in the former.
Due to these observations
it requires to reanalyse 
 $\hat{q}$  in the context of anisotropic
plasmas.
 Momentum broadening in a homogenous
but locally anisotropic high-temperature system for a heavy quark induced by collisions 
has been discussed recently \cite{Romatschke:2006bb} (see also
\cite{Romatschke:2004au,Dumitru:2007rp}).
To include the finite temperature  dependence in the quark-medium interaction 
we shall  consider the thermal quark self-energy
\cite{LeBellac:1996}
 in the approximation where the hard quark of momentum $p$ remains on-shell 
at temperature $T=0$. Thus, Eq. (\ref{eq:basic}) reads,
\beq
\hat{q} = - \frac{g^2 C_F}{2 p^0}~{\rm Im}~ \int  \frac{d^4q}{(2\pi)^4}~
q_\perp^2 ~ 2 \pi \delta_{+}((p-q)^2) ( 1 + f(q^0)) {\rm Tr}~[{p\slashh}
\gamma^\mu 
({p\slashh} - {q\slashh}) \gamma^\nu] \Delta_{\mu \nu}~,\label{eq:basic2}
\eeq

\begin{figure}[htb]
\begin{center}
\includegraphics[width=6cm]{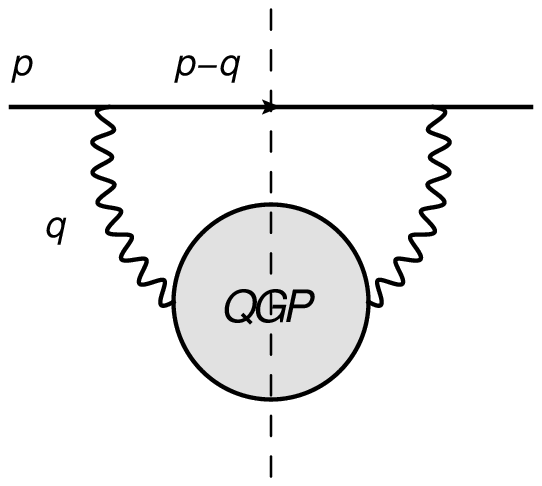}\label{fig1}\\
{\bf Figure 1:} Diagramatic representation of Eq.~(\ref{eq:basic2}).
\end{center}

\end{figure}

\vspace{0.2cm}

\noindent
(c.f. Fig.~1), where $ f(q^0) = ( \exp(q^0/T) -1)^{-1}$ is the Bose-Einstein distribution,
 and $\Delta_{\mu \nu}$ denotes the gluon propagator. In the eikonal
 approximation,
 when $q^0 << T$ and $q^0/p^0 \rightarrow 0$, one may simplify further Eq.~(\ref{eq:basic2}) :
\beq
\hat{q} = - g^2 C_F~{\rm Im}~ \int  \frac{d^4q}{(2\pi)^4}~
q_\perp^2 ~  \frac{2T}{q^0} 2 \pi \delta (q^0 - \vec{V} \cdot \vec{q}) 
~\frac{p\cdot  \Delta \cdot p}{(p^0)^2}~,
\label{finaleq}
\eeq
where $\vec{V}=\vec{p}/p^0$ is the quark velocity. In what follows, we shall
consider
 a massless quark, therefore we have $\vec{V}^2=1$.
Concerning kinematics: $\vec{V} \cdot \vec{q} = \vert \vec{q} \vert
\cos\theta_{pq} = q^0$,  $ q_\perp^2 =  \vert \vec{q} \vert^2
\sin^2 \theta_{pq} = \vert \vec{q} \vert^2 (1 - x^2)$ with $ x = q^0/ \vert \vec{q} \vert$.

As already pointed out in \cite{Romatschke:2006bb} this equation 
(\ref{finaleq}) tells us that all
the information about the medium,
either isotropic or anisotropic, is contained in the imaginary part of the
gluon propagator which is evaluated in the Hard-Thermal-Loop framework. 

\section{${\hat q}$ from Hard-Thermal-Loop}

In the Hard-Thermal-Loop approximation \cite{LeBellac:1996,Braaten:1989mz}
the inverse retarded gauge-field propagator in covariant gauge reads  
\begin{equation}
\left(\Delta^{-1}\right)^{\mu \nu}= -q^2 g^{\mu \nu} +q^\mu q^\nu
-\Pi^{\mu\nu}-\frac{1}{\lambda}q^\mu q^\nu ~.
\end{equation}
The gluon self-energy  is given by

\beq
\Pi^{\mu \nu}(q)= g^2 \int \frac{d^3 {\vec{p}}}{(2\pi)^3} \,
v^{\mu} \frac{\partial n({\vec{p}})}{\partial p^\beta}
 \left( g^{\nu \beta} -
\frac{v^{\nu} q^\beta}{q\cdot v + i \epsilon}\right) \; ,
\label{self} 
\eeq
where $v^{\mu} \equiv (1,{\vec{p}}/|{\vec{p}}|)$ is a light-like vector
describing the propagation of a plasma particle in space-time.

Following \cite{Romatschke:2003ms,Romatschke:2004jh},
 the phase space distribution function for an anisotropic plasma is taken as follows
\begin{equation}
n({\vec{p}}) = N(\xi) n_{\rm iso}\left(\sqrt{{\vec{p}}^2+\xi({\vec{p}}\cdot{\vec{
n}})^2} \right) ~.  \label{eq:f_aniso}
\end{equation}
Thus, $n({\vec{p}})$ is obtained from an isotropic distribution $n_{\rm
iso}(|\vec{p}|)$ by removing particles with a large momentum component
along $\vec{n}$. $N(\xi)=\sqrt{1+\xi}$ is a normalization factor which insures
that
  $\int d^3p\;n({\vec{p}})=\int d^3 p \;n_{\rm iso}({\vec{p}})$. In what
  follows,
 we shall omit this factor. Its effects on $\hat q$ will be discussed at the
 end of
 the section. \\
For evaluating Eq. (\ref{finaleq}) we choose  the reference frame in which the
 initial energetic (hard) quark propagates  along the $z$-axis, i.e, $\vec{V}  =
(0,0,1)$, whereas the beam nuclei collide along the $y$-axis, which is the
direction of anisotropy, denoted by the three-dimensional unit vector
${\bf{n}} = (0,1,0)$, namely,  $\vec{V}\perp \vec{n}$ which refered to a quark
produced 
at mid-rapidity . 

For an anisotropic plasma, the gluon propagator reads

\beq
\hat{p} \cdot \Delta \cdot \hat{p}=
\Delta_A ~\left[ 1 - x^2 - \frac{x^2 \qh^2}{1- \qh^2} \right] +
\Delta_G ~\left[x^2 (q^2 - \alpha -\gamma) + \frac{x^2 \qh^2}{1- \qh^2}
(\omega^2 - \beta) - 2 x^2 \qh \delta\right] ~,
\label{ pDp}
\eeq

where
\begin{equation}
\Delta^{-1}_{A} = (q^2-\alpha) ~,
\label{deltaa}
\end{equation}
and
\begin{equation}
\Delta^{-1}_{G} = (q^2-\alpha-\gamma)(\omega^2-\beta) + 
\delta^2 \vec{q}~^2 \tilde{n}^2~,
\label{deltag}
\end{equation}
with the transverse momentum component 
$\qh = \vec{q} \cdot \vec{n}/|\vec{q}| = q_y/|\vec{q}|$
into the direction of anisotropy. $\omega=q\cdot u$ where $u^{\mu}$ is the
heat-bath vector,
 which in the local rest frame
is given by $u^{\mu}=(1,0,0,0)$\\
The functions $\alpha, \beta, \gamma$ and $\delta$,
are obtained from Eq.~(\ref{self}). 
Explicit expressions maybe found e.g. in
 \cite{Romatschke:2003ms,Romatschke:2006bb}.

The contribution of $\Delta_A$ to $\hat{q}$ of Eq.~(\ref{finaleq})
is considered, which even to LL accuracy shows the possible presence
of the plasma instability. In performing the $|\vec{q}|= q-$integration,
 the contribution at LL accuracy reads

\beq
\hat{q}_A = - \frac{g^2 C_F T}{4 \pi^3}
\int d\Omega_q \frac{1 - x^2}{x} \left[ 1 - x^2 - \frac{x^2 \qh^2}{1- \qh^2} \right]
~I(x, \alpha) ~,
\label{qA}
\eeq
with
\beq
I(x, \alpha) \simeq  \frac{{\rm{Im}}\alpha}{ (1-x^2)^2} ~
\left[\frac{1}{2} \ln{\frac{T}{m_D}} + \frac{\pi}{2} \frac{{\rm{Re}}\alpha}{{\rm{Im}}\alpha}
\Theta(- {\rm{Re}}\alpha) \right] ~,
\label{Ising}
\eeq
where 
\beq
{\rm{Im}} \alpha \simeq - \frac{\pi}{4}~ x (1 - x^2) m_D^2~
\left\{ 1 + \frac{\xi}{2} [3 \qh^2 - 1 -x^2( 5 \qh^2 - 1)]
\right\}~,
\label{IMa}
\eeq 
and 
\beq
{\rm{Re}} \alpha \simeq  - \frac{1}{3} \xi \qh^2 m_D^2 ~,
\label{real}
\eeq
with $m_D$ the isotropic Debye mass. 
The second term in Eq.~(\ref{Ising}) has to be kept, because it reflects a singularity
for ${\rm{Im}}\alpha \propto x \rightarrow 0$ due to the anisotropy $\xi > 0$
that
 reflects the plasma instabilities present in an  anisotropic plasma. 

At small anisotropy, i.e., $\xi\ll 1$, analytic calculations are possible. 
Following \cite{Romatschke:2006bb} the contribution from
the first term in Eq.~(\ref{Ising}) to $\hat{q}_A$ is denoted as regular.
 After performing the angular 
integrations in Eq.~(\ref{qA})
it leads at LL order with $T >>m_D$ to

\beq
\hat{q}_A^{reg} = \frac{g^2 C_F m_D^2 T}{8 \pi} \ln \frac{T}{m_D}
~(1 + O(\xi^2))~,
\label{qhxi}
\eeq
with no contribution at first order in $\xi$.
 Applying the same procedure for the second term $\Delta_G$ we obtain 

\beq
\hat{q}_G^{reg} = \frac{3 g^2 C_F m_D^2 T}{8 \pi}
~\ln \frac{T}{m_D}~(1 + O(\xi^2)).
\label{qhg}
\eeq
with no singular contribution at this order.\\
 
Summing the two terms Eqs.~(\ref{qhxi}) and (\ref{qhg})
the LL transport coefficient in the limit of small $\xi$ up to $O(\xi)$
becomes
\beq
\hat{q}_{anisio}^{reg} = \hat{q}_A^{reg} +  \hat{q}_G^{reg} =
\frac{g^2 C_F m_D^2 T}{2 \pi}~ \ln\frac{T}{m_D}~ (1 + O(\xi^2)),
\label{regqhat}
\eeq
which reproduces  ${\hat q}$ in an isotropic plasma ($\xi=0$).  

Next the anomalous contribution \cite{Romatschke:2006bb} due to the second
term of Eq.~(\ref{Ising}) is evaluated. In LL order only the behavior
for $x \rightarrow 0$ is relevant. With Eq.~(\ref{real}) it gives
\beq
\hat{q}_A^{anom} \simeq  \frac{g^2 C_F m_D^2 T}{24 \pi^2}
\xi~ \int d\Omega_q \frac{\qh^2}{x} ~,
\label{anomq}
\eeq 
inducing a logarithmic singularity with
 $x = \cos \theta_{pq}$.
The contribution to $\hat{q}_G^{anom}$ starts at $O(\xi^2)$.
We note that the anomalous part (\ref{anomq}), that is the first 
correction produced by the anisotropy is positive.
 Therefore, the transport coefficient is enhanced by the anisotropy. 

To cut the singularity we suggest three possibilities: 

(i) One may 
follow the detailed and plausible 
arguments given in \cite{Romatschke:2006bb}
that this soft singularity is screened by
$O(g^3)$ terms in the gluon propagator, i.e. beyond the HTL approximation
under discussion. It leads to the replacement of  ${\rm{Im}} \alpha$
in the  second term in the denominator of Eq.~(\ref{Ising}) by
${\rm{Im}} \alpha  \sim x \rightarrow x + c g$, i.e. it is suggestive to cut the
singularity in (\ref{anomq}) by
\beq
\xi \int \frac{d x}{x} \rightarrow 2 \xi \int_0 \frac{d x}{x + cg}
\sim 2 \xi \ln\frac{1}{g} \sim 2 \xi \ln \frac{T}{m_D} ~.
\label{save}
\eeq
This way a finite result is obtained,
\beq
\hat{q}_A^{anom} \simeq  \frac{g^2 C_F m_D^2 T}{2 \pi}
\ln \frac{T}{m_D} ~ \frac{\xi}{6} ~,
\label{anomf}
\eeq 
which  shows a positive, but weak dependence on $\xi$ as a sign of the
anisotropy for $\xi > 0$..

(ii) The origin of the $1/x$ singularity is traced back to the Bose-Einstein
distribution $f(q^0) \sim T/q^0$ in Eq.~(\ref{eq:basic2}). Pragmatically,
in the anisotropic case, this behavior could be modified by
$q^0 \rightarrow \sqrt{(q^0)^2 + \xi (\vec{n} \cdot \vec{q})^2}
= q~\sqrt{x^2 + \xi \qh^2}, ~ q^0 > 0$. On mass-shell this replacement gives
the distribution in  Eq.~(\ref{eq:f_aniso}),  and  leads  to
\beq
\xi \int \frac{d x}{x} \rightarrow 2 \xi \ln \frac{1}{\xi} ~.
\label{save2}
\eeq

(iii) To form the anisotropic configuration in momentum space 
a characteristic time scale is present of the order
$\tau_c \sim O(1/g \xi T)$, for not to large $\xi$.
It is then natural to cut the energies of the constituents in the heat bath
by $|q^0| \ge 1/\tau_c > g \xi T$, and
\beq
\xi \int \frac{d x}{x} \rightarrow 2 \xi \ln \frac{1}{g \xi} ~.
\label{save3}
\eeq

In summary all three options Eqs.~(\ref{save} - \ref{save3}) lead to a positive
contribution of $O(\xi)$ at LL order to  $\hat{q}_{aniso}$. Also, if we take
into account
 the normalization factor $N(\xi)$ one gets an additional enhancement of $\hat
 q$
 by a factor $\sqrt{1+\xi}\simeq 1+\frac{\xi}{2}$ at small $\xi$.

\section{Conclusion}
To summarize, the transport coefficient for an anisotropic
 plasma is shown to be
  larger than that in an isotropic one, at least at small anisotropy for
 which
 the calculations are performed. However, a detailed quantitative study, 
for larger anisotropy, is needed to get an overall estimate of anisotropy
 effects
 on the jet-quenching parameter and for possible phenomenological applications
 at RHIC and LHC. Indeed, a better handling of the theoretical value of the
 transport
 coefficient can be essential to distinguish the various phenomenological
 models.
 Note that similar conclusions have been reached
 independently in 
 \cite{Dumitru:2007rp,Asakawa:2006jn,Muller:2007rs,Majumder:2009cf}.

\end{document}